\documentclass[12pt]{article}
\usepackage{epsfig,amsfonts,amsthm}
\usepackage{amsmath,amssymb}
\usepackage{array}
\usepackage{cite}
\usepackage{mathrsfs}
\usepackage{dsfont}
\usepackage[title]{appendix}
\usepackage{cancel}

\newcommand{\Et}{\cancel{E}_{T}}
\newcommand{\be}{\begin{equation}}
\newcommand{\ee}{\end{equation}}
\newcommand{\bea}{\begin{eqnarray}}
\newcommand{\eea}{\end{eqnarray}}

\newcommand{\doublet}[2]{ \left( \begin{array}{c}#1 \\ #2 \end{array}\right) }

\newcommand{\GeV}{{\ensuremath\rm \; GeV}}

\def\lsim{\mathrel{\rlap{\lower4pt\hbox{\hskip1pt$\sim$}}
    \raise1pt\hbox{$<$}}}         %less than or approx. symbol
\def\gsim{\mathrel{\rlap{\lower4pt\hbox{\hskip1pt$\sim$}}
    \raise1pt\hbox{$>$}}}         %greater than or approx. symbol

\newcommand{\bt}{\begin{tabular}}
\newcommand{\et}{\end{tabular}}

\usepackage{color}
\definecolor{myorange}{rgb}{1,0.5,0}

\usepackage[compat=1.0.0]{tikz-feynman}
 \tikzfeynmanset{compat=1.0.0} 
\usepackage{tikz}
\usetikzlibrary{decorations.pathmorphing,decorations.markings}

\usepackage{graphicx}%Täl saa laitettuu kuvii. Esim. ps filei. Kun haluu laittaa .epsi tai .eps tiedostoja, niin laittaa eka \includegraphics{nimi.eps}. Sit kääntää pdflatex:lla ja kuva tulee. 

\tikzset{
photon/.style={decorate, decoration={snake,amplitude=2pt, segment length=5pt}, draw=black},
particle/.style={draw=black, postaction={decorate}, decoration={markings,mark=at position .5 with {\arrow[draw=black]{>}}}},
antiparticle/.style={draw=black, postaction={decorate}, decoration={markings,mark=at position .5 with {\arrowreversed[draw=black]{>}}}},
gluon/.style={decorate, draw=black, decoration={coil,amplitude=4pt, segment length=5pt}},
goldstone/.style={draw=green,postaction={decorate},decoration={markings,mark=at position .5 with {\arrow[draw=blue]{>}}}}
}

\begin{document}

\title{\hfill ~\\[-30mm]
\begin{footnotesize}
\hspace{100mm}
DIAS-STP-22-01\\
\end{footnotesize}
\vspace{3mm}
                  \textbf{Complementary collider and astrophysical probes of multi-component Dark Matter}        }
%\date{}

\author{\\[-5mm]
J. Hern\'andez-S\'anchez\footnote{E-mail: {\tt jaime.hernandez@correo.buap.mx}}$^{~1}$,\ 
V.  Keus\footnote{E-mail: {\tt venus@stp.dias.ie}}$^{~2,3,4}$,
S.  Moretti\footnote{E-mail: {\tt stefano@soton.ac.uk, stefano.moretti@physics.uu.se}} $^{4,5,6}$,
%D.  Rojas\footnote{E-mail: {\tt drojas@ifuap.buap.mx}} $^{1,3}$,\  
D.  Soko\l{}owska\footnote{E-mail: {\tt dsokolowska@iip.ufrn.br}} $^{7,8}$
\\
\emph{\small $^1$ Instituto de F\'isica and Facultad de Ciencias de la Electr\'onica,}\\  
\emph{\small 
Benem\'erita Universidad Aut\'onoma de Puebla,
Apdo. Postal 542, C.P. 72570 Puebla, M\'exico,}\\
\textit{\small  $^2$  Dublin Institute for Advanced Studies, School of Theoretical Physics,}\\
\textit{\small 10 Burlington road, Dublin, D04 C932, Ireland}\\
\emph{\small $^3$ Department of Physics and Helsinki Institute of Physics,}\\
\emph{\small Gustaf Hallstromin katu 2, FIN-00014 University of Helsinki, Finland}\\
\emph{\small $^4$ School of Physics and Astronomy, University of Southampton,}\\
\emph{\small Southampton, SO17 1BJ, United Kingdom}\\
\emph{\small  $^5$ Particle Physics Department, Rutherford Appleton Laboratory,}\\
\emph{\small Chilton, Didcot, Oxon OX11 0QX, United Kingdom}\\
\emph{\small $^6$ Department of Physics and Astronomy, Uppsala University,}\\
\emph{\small Box 516, SE-751 20 Uppsala, Sweden}\\
\emph{\small  $^7$ University of Warsaw, Faculty of Physics, Pasteura 5, 02-093 Warsaw, Poland.}\\
  \emph{\small  $^8$ International Institute of Physics, Universidade Federal do Rio Grande do Norte,}\\
\emph{\small Campus Universitario, Lagoa Nova, Natal-RN 59078-970, Brazil}\\[1mm]
  }

\date{\vspace{-4ex}}

\maketitle

\begin{abstract}
\noindent
{We study a new physics scenario with two inert and one active scalar doublets, hence a 3-Higgs Doublet Model (3HDM). We impose a $Z_2 \times Z'_2$ symmetry onto such a 3HDM with one inert doublet odd under the $Z_2$ transformation and the other odd under the $Z'_2$ one. Such a construction leads to a two-component Dark Matter (DM) model. It has been shown that, when there is a sufficient mass difference between the two DM candidates, it is possible to probe the light DM candidate in the nuclear recoil energy in direct detection experiments and the heavy DM component in the photon flux in indirect detection experiments.
With the DM masses at the electroweak scale, we show that, independently of astrophysical probes, this model feature can be tested at the Large Hadron Collider via scalar cascade decays in $2\ell + \cancel{E}_T$ final states.
We study several observable distributions whose shapes hint at the presence of the two different DM candidates.
%In particular the processes $q \bar{q} \to Z^* \to H_{1,2} A_{1,2} \to H_{1,2} H_{1,2} Z^* \to H_{1,2} H_{1,2} l \bar{l}$ which shows up as $2l + \cancel{E}_T$, and the process $gg \to h \to A_{1,2}A_{1,2} \to Z^* H_{1,2} \, Z^* H_{1,2}$ which shows up as $4l + \cancel{E}_T$. 
%The mass splitting of the two DM candidates, $m_{H_2}-m_{H_1}$, is larger than the MET resolution at the LHC searches.
} 
 \end{abstract}
\thispagestyle{empty}
\vfill
\newpage
\setcounter{page}{1}

%\tableofcontents

%(To be added to the main draft)\\
%Question: Can experiments notice the existence of 2 DM candidates?\\
%Possible tricks: \\
%
%1) The plot of the count of events vs. the recoil energy has a different shape for the 
%2 DM candidates when the mass difference is large. Can this be distinguished in experiment? Is this trick not only possible useful to detect more than one DM candidate, but maybe to detect the existence of the other inert states?\\ 
%
%1.1) Can we separate elastic scattering from inelastic scattering? In this Reference \cite{Batell2011} they claim that, after distinguishing a DM event in DD experiments, one can then analise a difference between elastic and inelastic scattering events, from where the presence of two DM candidates could be noted.
%
%2) In the case of indirect detection, we can use for example the excess of photons observed by FermiLAT in the signature coming from the galactic center. If one DM candidate is not sufficient to account for these signature, we may prove that the contributions of two DM candidates adds up conveniently to match the observed excess.  See figure \ref{gammaflux}.
%Actually, the positrons excess is also interesting, an according to Reference \cite{Queiroz2020}, the signal can be explained with 2 DM candidates with very large mass splitting.\\
%
%3) If having DM candidates of opposite CP charge, can that help to test in experiments the existence of both candidates? See Reference \cite{DelleRose2018}.

\newpage

\section{Introduction}

The Standard Model (SM) of particle physics has been extensively tested and is in great agreement with experiment with its last particle – the Higgs boson $h$ – discovered in 2012 with a mass of $\approx$ 125 GeV by the ATLAS and CMS experiments at the CERN Large Hadron Collider (LHC) \cite{Aad:2012tfa,Chatrchyan:2012ufa}.  
Although the properties of the observed state are in agreement with those of the SM Higgs boson, it is entirely possible that it is one member of an extended scalar sector.

In fact, although in agreement with experiment, the SM is understood to be incomplete with 
one of its shortcomings being the lack of a viable Dark Matter (DM) candidate in its particle content.
The standard cosmological $\Lambda$CDM Model \cite{Ade:2015xua} requires DM to be a particle stable on cosmological time scales, cold (i.e., non-relativistic at the onset of galaxy formation), non-baryonic, neutral and weakly interacting:  such a state does not exist in the SM. 
Within Beyond the SM (BSM) frameworks, many such candidates exist, with the most well-studied being the Weakly Interacting Massive Particles (WIMPs) \cite{Jungman:1995df, Bertone:2004pz, Bergstrom:2000pn} with masses between a few GeV and a few TeV.
WIMPs are usually stable due to the conservation of a discrete symmetry, such as scalar DM candidates in non-minimal Higgs frameworks which are stabilised by the conserved discrete symmetry of the scalar potential, see, e.g.,  \cite{McDonald:1993ex,Burgess:2000yq,Deshpande:1977rw, Ma:2006km,Belanger:2012zr,Barbieri:2006dq,LopezHonorez:2006gr,Ivanov:2012hc}.

One of the simplest BSM scenarios which provides a scalar DM candidate is the Inert Doublet Model (IDM) \cite{Deshpande:1977rw}, which contains one inert doublet plus one active (Higgs) doublet, hence also known as I(1+1)HDM. 
In this model, which has been studied extensively in the literature (see, e.g., \cite{Ma:2006km,Barbieri:2006dq,LopezHonorez:2006gr}), the additional $SU(2)_W$ scalar doublet has the same SM quantum numbers as the SM Higgs doublet. 
One of the possible vacuum alignments in this model is $(v,0)$ wherein the {active} (or Higgs)   doublet acquires a non-zero Vacuum Expectation Value (VEV) while the {inert} (or dark) doublet does not develop a VEV and therefore does not take part in  electroweak symmetry breaking. 
The Lagrangian and the vacuum are symmetric under a $Z_2$ group under which only the inert doublet is odd.
Due to the conservation of this symmetry, the inert doublet provides a stable DM candidate: the lightest neutral $Z_2$-odd particle.

In the I(1+1)HDM, for masses of DM smaller than $m_h/2$, the dark sector communicates with the SM mainly through the Higgs boson exchange which is a characteristic of all Higgs-portal models \cite{Patt:2006fw,Chu:2011be,Queiroz:2014yna}. 
Such a set-up leads to the DM-Higgs coupling, $g_{{\rm DM}h}$, dictating the DM annihilation rate $\langle\sigma v\rangle$, the DM-nucleon scattering cross-section $\sigma_{\rm DM-N}$ and the Higgs invisible decays. 
Simultaneous fulfilment of current experimental constraints for these three types of processes is a challenging task, as shown, e.g., in \cite{Mambrini:2011ik,Djouadi:2011aa,Djouadi:2012zc}. For heavier DM particles, the direct annihilation into pairs of gauge bosons usually results in a relic density below the observed value, with the exception of very heavy DM particles.
A possible solution to this problem is breaking the simple relation between the annihilation rate and  direct detection cross-section by introducing coannihilation processes between DM and other dark particles which are close in mass. 
Coannihilation processes (through constructive and/or destructive interference) can decrease or increase the effective annihilation cross-section, which in turn change the DM relic density value.
In the I(1+1)HDM, the DM candidate could potentially coannihilate with the neutral or charged $Z_2$-odd particles. However, in a vast region of the parameter space, this coannihilation is too efficient, which leads to a total relic density below the observed value \cite{Arina:2009um,Nezri:2009jd,Miao:2010rg,Gustafsson:2012aj,Arhrib:2012ia,Krawczyk:2013pea,Goudelis:2013uca,Arhrib:2013ela,Krawczyk:2015vka,Ilnicka:2015jba,Diaz:2015pyv,Modak:2015uda,Queiroz:2015utg,Garcia-Cely:2015khw,Hashemi:2016wup,Poulose:2016lvz,Alves:2016bib,Datta:2016nfz,Belyaev:2016lok,Belyaev:2018ext,Sokolowska:2019xhe,Kalinowski:2019cxe}.

In models with an extended inert sector, more coannihilation processes are available and lead to a much richer phenomenology, see, e.g.,  models with 
extra inert singlets \cite{Belanger:2012vp,Yaguna:2019cvp,Belanger:2020hyh}
or inert doublets \cite{Ivanov:2012hc} and, within the framework of 3-Higgs Doublet Models (3HDMs), the I(2+1)HDM of Refs. \cite{Keus:2014jha,Keus:2014isa,Keus:2015xya,Cordero-Cid:2016krd,Cordero:2017owj,Cordero-Cid:2018man,Keus:2019szx,Aranda:2019vda,Cordero-Cid:2020yba}.
Proposed by Weinberg in 1976 \cite{Weinberg:1976hu}, 3HDMs are very well motivated scenarios \cite{Ivanov:2012fp,Keus:2013hya} due to their implications for flavour physics, CP-violation, baryogenesis and inflation \cite{Keus:2021dti,Keus:2020ooy,Keus:2021ojv,Davoudiasl:2019lcg,deMedeirosVarzielas:2016rii,Hartmann:2014ppa,GonzalezFelipe:2014mcf,Joaquim:2014gba}. 

Here, we study an I(2+1)HDM framework symmetric under a $Z_2\times Z'_2$ group with one inert doublet odd under $Z_2$ and even under $Z'_2$, and the other inert doublet even under $Z_2$ and odd under $Z'_2$.
The lightest particle from each inert doublet is a viable DM candidate, yielding a two-component DM model.
In a previous study \cite{Hernandez-Sanchez:2020aop}, we showed that other dark particles from both doublets influence the thermal evolution and decoupling rate of DM particles and significantly impact the final relic abundance.
(A similar analysis was performed in the context of a supersymmetric model in \cite{Khalil:2020syr}.)
When there is a sufficient mass difference between the two DM candidates, which are both typically at the Electro-Weak (EW) scale, we showed that the light DM component can be probed by the nuclear recoil energy in direct detection experiments while the heavy DM  component appears through its contribution to the photon flux in indirect detection experiments. 

In this paper, quite independently of astrophysical probes, we study collider signatures of the I(2+1)HDM, namely, scalar cascade decays in $2 \ell + \cancel{E}_T$ final states at the LHC. Specifically, we analyse several observable distributions whose shapes hint at the presence of the two different DM candidates.
The remainder of the paper is organised as follows. 
In Section \ref{sec:intro}, we present the model, its scalar mass spectrum and discuss theoretical and experimental constraints on its  parameter space.
In Section \ref{sec:probes}, we discuss the experimental probes of the model and construct some Benchmark Points (BPs).
In Section \ref{sec:results}, we present our results, and in Section \ref{sec:conclusion}, we conclude.

\section{Potential, mass spectrum and constraints}
\label{sec:intro}
\subsection{The scalar potential}

The most general $Z_2\times Z'_2$ symmetric 3HDM potential has the following form \cite{Ivanov:2011ae,Keus:2013hya}:  

\bea
\label{potential}
V &=& V_0+V_{Z_2 \times Z'_2},\\[1mm]
V_0 &=& - \mu^2_{1} (\phi_1^\dagger \phi_1) -\mu^2_2 (\phi_2^\dagger \phi_2) - \mu^2_3(\phi_3^\dagger \phi_3) \nonumber
+ \lambda_{11} (\phi_1^\dagger \phi_1)^2+ \lambda_{22} (\phi_2^\dagger \phi_2)^2  + \lambda_{33} (\phi_3^\dagger \phi_3)^2 \nonumber\\
&& + \lambda_{12}  (\phi_1^\dagger \phi_1)(\phi_2^\dagger \phi_2)
 + \lambda_{23}  (\phi_2^\dagger \phi_2)(\phi_3^\dagger \phi_3) + \lambda_{31} (\phi_3^\dagger \phi_3)(\phi_1^\dagger \phi_1) \nonumber\\
&& + \lambda'_{12} (\phi_1^\dagger \phi_2)(\phi_2^\dagger \phi_1) 
 + \lambda'_{23} (\phi_2^\dagger \phi_3)(\phi_3^\dagger \phi_2) + \lambda'_{31} (\phi_3^\dagger \phi_1)(\phi_1^\dagger \phi_3),  \nonumber\\[1mm]
V_{Z_2 \times Z'_2}&=&  \lambda_1 (\phi_1^\dagger \phi_2)^2 + \lambda_2(\phi_2^\dagger \phi_3)^2 + \lambda_3 (\phi_3^\dagger \phi_1)^2 + \mathrm{h.c.}, \nonumber 
\eea
where $V_0$ is invariant under any phase rotation while $V_{Z_2 \times Z'_2}$ ensures the symmetry under the $Z_2 \times Z'_2$ group generated by
\be 
g_{Z_2} = \mathrm{diag}(-1,1,1)\, , \qquad
g_{Z'_2} = \mathrm{diag}(1,-1,1) \,.
\ee
Under this charge assignment, all SM fields, including the Higgs doublet $\phi_3$, are even under both $Z_2$ and $Z_2'$.  The additional doublets, $\phi_1$ and $\phi_2$ are odd under $Z_2$ and $Z_2'$, respectively.   In this paper, we assume that all parameters in the potential are real, therefore, we do not introduce any explicit CP-violation in the scalar sector. In particular, we do not consider the possible effects of dark CP-violation, a feature that was introduced for the first time in \cite{Cordero-Cid:2016krd}
and further studied in 
\cite{Keus:2016orl,Cordero:2017owj,Cordero-Cid:2018man,Cordero-Cid:2020yba},
which could indeed arise in an extended dark sector. However, there is still the possibility of spontaneous breaking of the CP symmetry in the {active} sector for particular choices of parameters, as discussed in \cite{Hernandez-Sanchez:2020aop}. This choice of vacuum is not considered in this paper as it cannot lead to the  appearance of two DM candidates in the model.

The Yukawa interactions are set to ``Type-I'' interactions, i.e.,  only the third doublet, $\phi_3$, will couple to fermions:
\be
\mathcal{L}_{Y} = \Gamma^u_{mn} \bar{q}_{m,L} \tilde{\phi}_3 u_{n,R} + \Gamma^d_{mn} \bar{q}_{m,L} \phi_3 d_{n,R}
 +  \Gamma^e_{mn} \bar{l}_{m,L} \phi_3 e_{n,R} + \Gamma^{\nu}_{mn} \bar{l}_{m,L} \tilde{\phi}_3 {\nu}_{n,R} + \mathrm{h.c.}  \label{yukawa}
\ee
Following the $Z_2 \times Z_2'$ charge assignment, this choice of Yukawa interaction is the only one which will not lead to breaking of the imposed discrete symmetries. This also ensures that there are no Flavour Changing Neutral Currents (FCNCs), since fields from the doublets which do not develop VEVs, $\phi_1$ and $\phi_2$,
will not couple to fermions.

\subsection{Mass spectrum and physical parameters}
\label{sec:mass}
 
The \texttt{2Inert} vacuum state, i.e., the one with two automatically stable inert particles,  has the alignment $(0,0,v)$ in which the composition of the doublets are
\be 
\phi_1= \doublet{$\begin{scriptsize}$ H^+_1 $\end{scriptsize}$}{\frac{H_1+iA_1}{\sqrt{2}}} ,\quad 
\phi_2= \doublet{$\begin{scriptsize}$ H^+_2 $\end{scriptsize}$}{\frac{H_2+iA_2}{\sqrt{2}}} , \quad 
\phi_3= \doublet{$\begin{scriptsize}$ H^+_3 $\end{scriptsize}$}{\frac{v+h+iA^0_3}{\sqrt{2}}} \, ,
\label{vac-inert}
\ee
with the extremum condition for this state reading as
\be 
v^2=\frac{\mu_3^2}{\lambda_{33}}\,.
 \label{inertex}
\ee
The third doublet, $\phi_3$, plays the role of the SM Higgs doublet, with the Higgs particle $h$ having, by construction, tree-level interactions with gauge bosons and fermions identical to those of the SM Higgs boson. Its mass is fixed through the tadpole conditions to be
\be
m^2_h= 2\mu_3^2 = 2 v^2 \lambda_{33}
\ee
and the ${A^0}_3$ and ${H}^\pm_3$ states are the would-be Goldstone bosons. 

The two inert doublets, $\phi_1$ and $\phi_2$, provide two DM candidates. Each doublet consists of two neutral particles\footnote{As it is the case in multi-scalar models with unbroken $Z_2$ symmetries, the  inert scalars $H_i$ and $A_i$ have opposite CP parity, as evident from their gauge interactions, however, it is not possible to establish their definite CP properties, as they do not couple to fermions.}, $H_i$ and $A_i$, and one charged particle $H^\pm_i$ with $i=1,2$. For the two generations of inert scalars the mass spectrum is as follows:
\bea
&& m^2_{H_1}= -\mu^2_1  +\frac{1}{2}(\lambda_{31}+\lambda'_{31} +2\lambda_3 )v^2 \equiv -\mu_1^2 + \Lambda_3 v^2, \label{massmh1}\\
&& m^2_{A_1}= -\mu^2_1  +\frac{1}{2}(\lambda_{31}+\lambda'_{31} -2\lambda_3)v^2 \equiv -\mu_1^2 + \bar{\Lambda}_3 v^2,\\
&& m^2_{H^\pm_1}= -\mu^2_1 +\frac{1}{2}\lambda_{31}v^2
\eea
and
\bea 
&& m^2_{H_2}= -\mu^2_2  +\frac{1}{2}(\lambda_{23}+\lambda'_{23} +2\lambda_2)v^2\equiv -\mu_2^2 + \Lambda_2 v^2, \label{massmh2}\\
&& m^2_{A_2}= -\mu^2_2  +\frac{1}{2}(\lambda_{23}+\lambda'_{23} -2\lambda_2)v^2\equiv -\mu_2^2 + \bar{\Lambda}_2 v^2,\\
&& m^2_{H^\pm_2}= -\mu^2_2 +\frac{1}{2}\lambda_{23} v^2. \label{massmhc2}
\eea

Parameters of the potential can be rephrased in terms of  physical observables, such as masses and couplings. The tree-level SM couplings in the gauge and fermionic sectors follow exactly the SM definitions. The relevant parameters arising from the extended scalar sector are: (i) masses of inert particles and the Higgs-DM couplings, which represent parameters from the visible sector; (ii) self-interaction parameters, which describe interaction within the dark sector. The full list is:
\be
v^2, m_h^2, m^2_{H_1}, m^2_{H_2}, m^2_{A_1}, m^2_{A_2}, m^2_{H^\pm_1}, m^2_{H^\pm_2}, \Lambda_2, \Lambda_3, \Lambda_1, \lambda_{11}, \lambda_{22}, \lambda'_{12}, \lambda_{12} \label{physpar}.
\ee
The self-couplings $\lambda_{11}, \lambda_{22}, \lambda'_{12}, \lambda_{12}$ correspond exactly to the terms in Eq.  (\ref{potential}), while the relations between remaining parameters and our chosen physical basis are as follows:
\bea
&& \mu_1^2=-m_{H_1}^2+\Lambda_{3}v^2, \\
&& \lambda_{3}=(m_{H_1}^2-m_{A_1}^2)/(2v^2), \\
&& \lambda_{31}'=(m_{H_1}^2+m_{A_1}^2-2m_{H^\pm_1}^2)/v^2, \\
&& \lambda_{31}=2\Lambda_{3}-2\lambda_{3}-\lambda_{31}',\\
&& \mu_2^2=-m_{H_2}^2+\Lambda_{2}v^2, \\
&& \lambda_{2}=(m_{H_2}^2-m_{A_2}^2)/(2v^2),\\
&& \lambda_{23}'=(m_{H_2}^2+m_{A_2}^2-2m_{H^\pm_2}^2)/v^2, \\
&& \lambda_{23}=2\Lambda_{2}-2\lambda_{2}-\lambda_{23}',\\
&& \lambda_1 = 2\Lambda_1 - (\lambda_{12}+\lambda'_{12}).
\eea

In principle, any particle among $(H_i, A_i, H^\pm_i)$ can be the lightest. Here, we dismiss the possibility of $H^\pm_i$ being the lightest, as it would mean that DM candidate is a charged particle. Choosing between $H_1$ and $A_1$ (or $H_2$ and $A_2$) is related only to a change of the sign of the quartic parameter $\lambda_{3}$ ($\lambda_2$) and has no impact on the ensuing phenomenology. Therefore, we will choose  $m_{H_i} < m_{A_i}, m_{H^\pm_i}$, which leads to the following relations between the parameters:
\be
\lambda_2<0, \;\; \lambda_3<0, \;\; \lambda_{31}'+ 2\lambda_3<0, \;\; \lambda_{23}'+ 2\lambda_2<0. \label{Hlightest}
\ee
Notice that, unlike many $Z_N$ symmetric models, the two lightest states from two doublets are automatically stable, regardless of their mass hierarchy, as they are stabilised by different $Z_2$ symmetries.

\subsection{Theoretical and experimental constraints}
The parameters of the potential $V$ are subject to a number of  theoretical and experimental constraints (described in detail  in 
\cite{Hernandez-Sanchez:2020aop}). Below, we summarise the constraints imposed on the model to ensure that all proposed BPs  are in agreement with current theoretical and experimental knowledge.

\paragraph{Stability of the potential} 
For the potential to be bounded from below (i.e., having a stable vacuum) the following conditions are required 
\cite{Grzadkowski:2009bt}:
\bea
&& \lambda_{ii}>0, \quad i =1,2,3, \label{positivity1} \\
&& \lambda_x > - 2 \sqrt{\lambda_{11} \lambda_{22}}, \quad \lambda_y > - 2 \sqrt{\lambda_{11} \lambda_{33}}, \quad \lambda_z > - 2 \sqrt{\lambda_{22} \lambda_{33}}, \label{positivity2}\\
&&\left\lbrace  \begin{array}{l} 
\sqrt{\lambda_{33}} \lambda_x + \sqrt{\lambda_{11}} \lambda_z+\sqrt{\lambda_{22}} \lambda_y \geq 0\\
\quad \textrm{or}\\[1mm]
\lambda_{33} \lambda_x^2 + \lambda_{11} \lambda_z^2+\lambda_{22} \lambda_y^2 -\lambda_{11} \lambda_{22} \lambda_{33} - 2 \lambda_x \lambda_y \lambda_z <0,
\end{array}\right.
\label{positivity3}
\eea
where 
\bea
\lambda_x = \lambda_{12}+\textrm{min}(0,\lambda_{12}'-2|\lambda_1|),\\
\lambda_y = \lambda_{31}+\textrm{min}(0,\lambda_{31}'-2|\lambda_3|),\\
\lambda_z = \lambda_{23}+\textrm{min}(0,\lambda_{23}'-2|\lambda_2|).
\eea
As noted in \cite{Faro:2019vcd},  these conditions are in fact sufficient but
not necessary, as it is possible to construct examples of this model in which the potential is bounded from below, but which violate conditions
 (\ref{positivity1})--(\ref{positivity3}).  We do not explore such a region of parameter space in this work. 

\paragraph{Global minimum condition}
For $(0,0,v)$ to be a local minimum, all mass-squared values have to be positive, and, for it to be a global minimum, i.e., the true vacuum, its energy, $\mathcal{V}_{\texttt{2Inert}}$, has to be lower than the energy of any other possible minima, $\mathcal{V}_{\texttt{X}}$, that exist at the same time (see discussion in \cite{Hernandez-Sanchez:2020aop}). Following the chosen mass order and resulting relations in Eq.~(\ref{Hlightest}) we arrive at the following conditions:
\bea
 \textrm{Local minimum if:}&& \left\lbrace \begin{array}{l}
v^2={\mu_3^2}/{\lambda_{33}} > 0\\[1mm]
\Lambda_2 > \mu^2_2/v^2  \\[1mm]
\Lambda_3 > \mu^2_1/v^2. \end{array} \right. 
\hspace{3cm}
\label{inert-loc}
\\
\textrm{Global minimum if, in addition:}&& \mathcal{V}_{\texttt{2Inert}}= -\frac{\mu_3^4}{4\lambda_{33}} < \mathcal{V}_{\texttt{X}}. 
\hspace{3cm}
\label{inert-glob}
\eea
\paragraph{Perturbative unitarity} We require that the scalar $2 \to 2$ scattering matrix is unitary, i.e., the absolute values of all eigenvalues of such a matrix for Goldstones, Higgs and dark states with specific hypercharge and isospin should be smaller than $8 \pi$. Furthermore, all quartic scalar couplings should be perturbative, i.e., $\lambda_i \leq 4 \pi$.

\paragraph{EW Precision Observables (EWPOs)} We demand a 2$\sigma$, i.e., 95\% Confidence Level (CL) agreement with EWPOs which are  parametrised through the EW oblique parameters $S,T,U$. Assuming an SM Higgs boson mass of $m_h$ = 125 GeV, the central values of the oblique parameters are given by~\cite{Baak:2014ora}:
\be 
\hat{S} = 0.05 \pm 0.11 ,\qquad \hat{T} = 0.09 \pm 0.13, \qquad \hat{U}=0.01\pm 0.11.
\label{eq:ewpt}
\ee
In the I(1+1)HDM these constraints impose a strict order on the masses of the inert particles, with two neutral dark scalars being lighter than the charged particle. Furthermore, mass splitting between heavier neutral scalar and charged scalar is limited to roughly 50 GeV. However, in the case of a $Z_2 \times Z_2'$ 3HDM,  these conclusions are no longer necessary. Cancellations between contributions to $S,T,U$ parameters from the two generations of dark particles may lead to a different mass orderings, where either of $A_i$ or $H^\pm_i$ is the heaviest, as well as to an increased mass splittings between these particles (for a detailed discussion, see \cite{Hernandez-Sanchez:2020aop}).

\paragraph{Collider searches for new physics} The presence of additional scalars, especially if they are sufficiently light, can influence properties of SM particles, e.g., their decay channels and widths. We forbid decays of EW gauge bosons into new scalars by enforcing:
\be 
\label{eq:gwgz}
m_{H_i}+m_{H^\pm_i}\,\geq\,m_W^\pm,~~ m_{A_i}+m_{H^\pm_i}\,\geq\,m_W^\pm,~~
\,m_{H_i}+m_{A_i}\,\geq\,m_Z,\,~~
2\,m_{H_i^\pm}\,\geq\,m_Z.
\ee
Furthermore, we adopt LEP 2 searches for supersymmetric particles re-interpreted for the I(1+1)HDM in order to exclude the region of masses where the following conditions are simultaneously satisfied \cite{Lundstrom:2008ai} ($i=1,2$):
\be 
\label{eq:leprec}
m_{A_i}\,\leq\,100\,\GeV,\,~~
m_{H_i}\,\leq\,80\,\GeV,\,\, ~~
\Delta m= |m_{A_i}-m_{H_i}|\,\geq\,8\,\GeV,
\ee
since this would lead to a visible di-jet or di-lepton signal.

The model also must agree with null results for additional neutral scalar searches at the LHC. 
As discused in \cite{Hernandez-Sanchez:2020aop}, current searches at the LHC for multi-lepton final states with missing transverse energy, $\cancel{E}_T$, are, in general, not sensitive enough to probe the parameter space of this model. This is mainly due to a relatively large cut on $\cancel{E}_T$ used in  experimental analyses, which results in a reduced sensitivity to probe the viable parameter space of the I(2+1)HDM scenario. Notice also that, as new charged particles are inert and hence do not couple to fermions, they are not subject to many constraints present in the 2HDM framework, e.g., flavour bounds on the charged scalar mass from $b\to s \gamma$, are not applicable here.

\paragraph{Charged scalar mass and lifetime}

We take a model independent lower estimate on the masses of all charged states: $m_{H^\pm_i} > 70$ GeV ($i=1,2$) \cite{Pierce:2007ut}. Furthermore, in this work we will not consider scenarios with possibly long-lived charged particles and, following \cite{Heisig:2018kfq}, we set the limit for a charged state lifetime to be $\tau\,\leq\,10^{-7}\,{\rm s}$ .

\paragraph{Higgs mass and signal strengths} The combined ATLAS and CMS result for the Higgs mass is~\cite{ATLAS:2015yey}:
\be
m_h = 125.09\pm 0.21 \textrm{ (stat.)} \pm 0.11 \textrm{ (syst.)} \; \GeV.
\ee
The Higgs particle detected at the LHC is  in excellent agreement with the SM predictions. 
By construction, the $h$ state in the \texttt{2Inert} vacuum in Eq.~(\ref{vac-inert}) is SM-like and its (tree-level) couplings to gluons, massive gauge bosons and fermions are identical to the SM values.
 
The Higgs total width can be modified through additional decays into light inert scalars, $S$, by contributing to the $h\to SS$ decay channel when $m_S \leq m_h/2$ as well as through modifications to decay channels already present in the SM, in particular, the $h \to \gamma \gamma$ decay. In this work we take the upper limit on the Higgs total decay width to be \cite{Sirunyan:2019twz}:
\be
\Gamma_{\rm tot} \leq 9.1 \; \textrm{MeV}.
\ee
The partial decay width $\Gamma(h\to \gamma\gamma)$ is modified with the respect to the SM through the presence of two charged inert scalars. In this work, we use the combined ATLAS and CMS limit for the signal strength \cite{Khachatryan:2016vau}:
\be
\mu_{\gamma \gamma} = 1.14^{+0.19}_{-0.018},
\ee
ensuring a 2$\sigma$ agreement with the observation.

The latest constraints on the Higgs invisible decays from CMS and ATLAS are \cite{Sirunyan:2018owy, Aaboud:2019rtt}:
\be
\textrm{BR}(h \to \textrm{ inv.}) < 0.19 \; (\textrm{CMS}), ~~ 0.26 \; (\textrm{ATLAS}).
\ee
These constraints significantly limit the allowed values of Higgs-inert couplings for light inert states.

\paragraph{DM constraints}
The total relic density is given by the sum of the contributions from both DM candidates $H_1$ and $H_2$,
\be 
\label{planck-relic}
\Omega_{T}h^2 = \Omega_{{H_1}}h^2 + \Omega_{{H_2}}h^2 \, ,
\ee
and is constrained by Planck data \cite{Aghanim:2018eyx} to be:
\be
\Omega_{\text{\rm DM}}h^2 = 0.1200 \pm 0.0012.
\label{PLANCK_lim}
\ee
The current strongest upper limit on the Spin-Independent (SI) scattering cross-section of DM particles off of nuclei,
$\sigma_{\rm DM-N}$, is provided by the XENON1T experiment and is relevant for all regions of DM mass \cite{Aprile:2018dbl}. Regarding indirect detection searches, for light DM particles annihilating into $bb$ or $\tau\tau$, the strongest constraints come from the Fermi-LAT satellite, ruling out the canonical cross-section 
$\langle \sigma v\rangle \approx 3\times 10^{-26}~{\rm cm}^3/{\rm s}$ for $m_{\rm DM} \lesssim 100 \mbox{ GeV}$ 
\cite{Ackermann:2015zua}.
For heavier DM candidates the PAMELA and Fermi-LAT experiments provide similar limits of $ \langle \sigma v\rangle \approx 10^{-25}~{\rm cm}^3/{\rm s}$ for  $m_{\rm DM}=200 \mbox{ GeV} $ in the $bb,\tau\tau$ or $WW$ channels \cite{Cirelli:2013hv}.

\section{Experimental probes of the model}
\label{sec:probes}

A detailed DM analysis of the $Z_2 \times Z'_2$ symmetric I(2+1)HDM was presented in \cite{Hernandez-Sanchez:2020aop}, with an emphasis on astrophysical
probes of it. Here we focus instead on the complementarity between astrophysical and collider tests of the model. 
In our present analysis, we use the \texttt{micrOMEGAs} package~\cite{Belanger:2006is} to calculate the relic density of the two DM candidates. During the analysis we follow the standard assumptions included in the code, namely: 1) particles within the dark sector are in thermal equilibrium;  2) they have the same kinetic temperature as that of the SM particle bath;  3) the number densities  of DM particles can differ from the equilibrium values  once their number density multiplied by their annihilation cross-section becomes too small to compete with the expansion rate of the Universe.  

The lightest particle in each family is a viable DM candidate. As discussed previously, without loss of generality, we can take $H_1$ from the first family and $H_2$ from the second family to be the respective DM candidates. However, other dark particles from both families have a significant impact on the final relic abundances of these two stable particles, as they influence the thermal evolution and decoupling rate of DM particles.

In the discussion that follows, $x_{a}$ denotes any relevant dark particle for a particular process from a respective dark sector. There are two\footnote{Due to the imposed symmetry there are no processes that would be classified as semi-annihilation, i.e., processes of the form $x_a x_b \to \textrm{SM } x_d$. } distinct classes of processes that can influence number densities of dark particles in each sector. 
In the first class there are (co)annihilation processes of the type:
\be
x_a x_a \to \textrm{ SM SM}.
\ee
In this class of processes, we have the standard DM annihilation, $H_i H_i \to  \textrm{ SM SM}$ with $i=1,2$, whose product depends mostly on the mass of the DM particles: we observe mostly Higgs-mediated annihilation into fermions for relatively light DM ($m_{\rm DM} \lesssim m_h/2$), while heavier DM particles annihilate predominantly into gauge boson pairs, either directly or through Higgs $s$-channel. 
In our calculations, we also include annihilation into virtual gauge bosons, as these processes have a significant impact on DM annihilation rates for medium DM masses ($m_h/2 \lesssim m_{\rm DM} \lesssim m_{W^\pm}$). 
Furthermore, if the mass difference between the DM candidate and other neutral or charged inert scalars from the same generation is small, then coannihilation channels such as $H_i A_i \to Z \to \textrm{SM SM}$ play an essential role. 
This is, in fact, the dominant class of processes for relatively high DM masses ($m_{\rm DM} \gtrsim 500$ GeV), exactly as it happens in the I(1+1)HDM. We would like to stress that, due to the $Z_2 \times Z'_2$ imposed symmetry in this class of processes, the two DM sectors are separated. 
There are no vertices that involve fields from two separate families, e.g., Higgs or gauge bosons couple only to a pair of inert particles from the same generation.

\begin{figure}[h!]
\begin{center}
\begin{tikzpicture}[thick,scale=0.75]
    \begin{feynman}
        \vertex (m) at ( 0, 0);
        \vertex (n) at (1.5,0);
        \vertex (a) at (-1.5,-1.5) {\(H_2\)};
        \vertex (b) at ( 3,-1.5) {\(H_1\)};
        \vertex (c) at (-1.5, 1.5) {\(H_2\)};
        \vertex (d) at ( 3, 1.5) {\(H_1\)};
        \vertex (e) at ( 1, -2) {\((a)\)};  
		\vertex (f) at ( 3.5, 0) {\(\propto \Lambda_2 \cdot \Lambda_3\)};  
        \diagram* {
            (a) -- [scalar] (m),
            (b) -- [scalar] (n),
            (c) -- [scalar] (m),
            (d) -- [scalar] (n),
			(n) -- [scalar, edge label'=\(h\)] (m),
        };
    \end{feynman}
\end{tikzpicture} 
\hspace{10mm}
\begin{tikzpicture}[thick,scale=0.75]
    \begin{feynman}
        \vertex (m) at ( 0, 0);
        \vertex (a) at (-1.5,-1.5) {\(H_2\)};
        \vertex (b) at ( 1.5,-1.5) {\(H_1\)};
        \vertex (c) at (-1.5, 1.5) {\(H_2\)};
        \vertex (d) at ( 1.5, 1.5) {\(H_1\)};
        \vertex (e) at ( 0, -2) {\((b)\)};        
		\vertex (f) at ( 2, 0) {\(\propto \Lambda_1\)};  
        \diagram* {
            (a) -- [scalar] (m) -- [scalar] (c),
            (b) -- [scalar] (m) -- [scalar] (d),
        };
    \end{feynman}
\end{tikzpicture}
\hspace{10mm}
\begin{tikzpicture}[thick,scale=0.75]
    \begin{feynman}
        \vertex (m) at ( 0, 0);
        \vertex (n) at (1.5,0);
        \vertex (a) at (-1.5,-1.5) {\(A_2\)};
        \vertex (b) at ( 3,-1.5) {\(H_1\)};
        \vertex (c) at (-1.5, 1.5) {\(H_2\)};
        \vertex (d) at ( 3, 1.5) {\(A_1\)};
        \vertex (e) at ( 1, -2) {\((c)\)};  
		\vertex (f) at ( 3, 0) {\(\propto g^2\)};  
        \diagram* {
            (a) -- [scalar] (m),
            (b) -- [scalar] (n),
            (c) -- [scalar] (m),
            (d) -- [scalar] (n),
			(n) -- [boson, edge label'=\(Z\)] (m),           };
    \end{feynman}
\end{tikzpicture} 
\caption{Example of DM conversion diagrams: (a) Higgs-mediated conversion of $H_2 H_2 \to H_1 H_1$, present always as long as $\Lambda_{2,3} \neq 0$; (b) direct DM conversion depending on the self-interaction parameter $\Lambda_1$; (c) $Z$-mediated conversion due to coannihilation channels.}
 \label{fig-conversion}
 \end{center}
\end{figure}
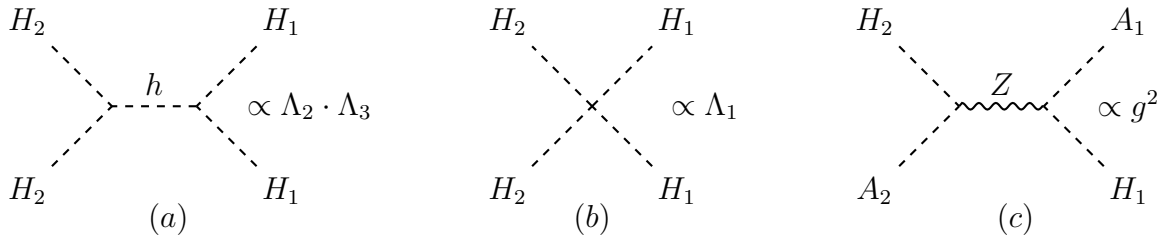

The second class of processes is DM conversion in which a pair of heavier scalars from one generation converts, either directly or through interaction with an SM particle, into a pair of dark particles from the other generation:
\be
x_a x_a \to x_b x_b \, .
\ee
Note that the conversion between two generations of DM particles occurs even if all self-interaction couplings are switched off. In Fig.~\ref{fig-conversion}, diagrams (a) and (b) represent the same initial/final state where a pair of $H_2$ particles are converted to a pair of $H_1$ particles through either Higgs-mediated or direct conversion. 
Even if the self-interaction parameter $\Lambda_1$ was set to zero, there would still be a non-zero contribution coming from diagram (a), as long as both particles couple to the Higgs boson ($\Lambda_{2,3}\neq 0$). 
We expect cancellations or enhancements depending on the relative sign of $\Lambda_2 \Lambda_3$ and $\Lambda_1$.
% in a similar vein to the effect observed in the standard annihilation into gauge boson pair. 
Furthermore, depending on the masses and couplings, we also need to take into account annihilation of heavier dark particles from the second generation directly into stable particles from the first generation, e.g., $A_2 A_2 \to h \to H_1 H_1$. All these processes are automatically included in our numerical analysis.

As discussed in Section \ref{sec:mass} in Eq.~(\ref{physpar}) we can parametrise the model by using masses of scalar particles and their couplings as follows.
\begin{enumerate}
\item 
The masses of  the inert particles, $m^2_{H_1}, m^2_{H_2}, m^2_{A_1}, m^2_{A_2}, m^2_{H^\pm_1}, m^2_{H^\pm_2}$ dictate the annihilation patterns of DM particles. Depending on the absolute values of masses, but also on mass splittings between particles, we can expect different dominant channels of annihilation, coannihilation and conversion. Furthermore, both absolute and relative values of these masses will result in different possible collider signatures.

\item 
The couplings of the DM particles to the Higgs boson, $\Lambda_2$ and $\Lambda_3$, govern not only DM annihilation and conversion but also influence possible invisible decays of the Higgs particle as well as direct and indirect detection of DM. 
In our numerical analysis we find that, in particular, the following vertices have significant impact on DM phenomenology:
\bea
g_{hH_1H_1} &=& 2\lambda_3 +\lambda_{31} + \lambda'_{31} = 2 \Lambda_3 \,, 
\\
g_{hH_2H_2} &=& 2\lambda_2 +\lambda_{23} + \lambda'_{23} = 2 \Lambda_2 \,, \\
g_{hA_1A_1} &=& -2\lambda_3 +\lambda_{31} + \lambda'_{31} = 2\Lambda_3 + 2(m^2_{A_1} - m^2_{H_1})/v^2  \,,  
\label{Eq:cubic-couplings}
\\
g_{hA_2A_2} &=&  -2\lambda_2 +\lambda_{23} + \lambda'_{23} = 2\Lambda_2 + 2(m^2_{A_2} - m^2_{H_2})/v^2 \,.
\eea

\item 
The self-couplings of dark particles, $\lambda_1, \lambda'_{12}, \lambda_{12}$ and $\lambda_{11}, \lambda_{22}$, play two different roles. The first set governs interactions between two different families and will have an observable impact on DM relic abundance through DM conversion processes. In particular, the following couplings have a crucial impact on DM phenomenology:
\bea
g_{H_1H_1H_2H_2} &=&  \phantom{-} 2\lambda_1 +\lambda_{12} + \lambda'_{12} =  4\Lambda_1 -(\lambda_{12} + \lambda'_{12}) \,,
\\
g_{A_1A_1H_2H_2}  = g_{A_2A_2H_1H_1} &=&  -4\lambda_1 +\lambda_{12} + \lambda'_{12} =-4\Lambda_1 + 2(\lambda_{12} + \lambda'_{12})  \,.
%\\
%\red{g_{A_1A_1H_2H_2}  = g_{A_2A_2H_1H_1} }
%&\red{=}& 
%\red{-2\lambda_1 +\lambda_{12} + \lambda'_{12} =-4\Lambda_1 + 3(\lambda_{12} + \lambda'_{12})} \nonumber
\eea
In turn, $\lambda_{11}$ and $\lambda_{22}$ do not directly contribute to any observable process and do not influence the DM abundance. They also have no impact on any observable collider processes. However, they have a fundamental impact on the range of other parameters through vacuum stability conditions.
\end{enumerate}

Our numerical analysis shows that, due to the existence of the conversion processes, the total DM relic density receives its dominant contribution from $H_1$  while the contribution from $H_2$ is of a few percent of the total DM relic density.

\subsection{BPs}
\label{sec:BPs}

In our previous analysis \cite{Hernandez-Sanchez:2020aop}, we showed that, as a consequence of both DM candidate masses being in the EW region, astrophysical tools could be used to probe different DM components in a complementary way. The lighter DM candidate, $H_1$, from which the total DM relic density receives dominant contribution, could have a detectable effect in the spectrum of the nuclear recoil energy measured at direct direction experiments (e.g., XENONnT/LZ or DARWIN) \cite{Aprile:2015uzo,Aalbers:2016jon,Baudis:2013qla}. 
Complementary to that, the heavier DM candidate, $H_2$, whose contribution to the total DM relic density is sub-dominant, may be detectable in its effect in enhancing the  photon flux coming from the galactic centre measured at indirect detection experiments (e.g., FermiLAT) \cite{Ackermann:2012qk}.

In the present paper, we focus on yet another complementary and independent probe of the two-component DM nature of this model, namely its collider signatures. 
We study  scalar cascade decays into the two different DM candidates, $H_1$ and $H_2$. 
Since the latter have comparable masses, we show that their presence is detectable simultaneously. Furthermore,  their mass difference leads to different shapes in the distributions of several observables such as $\cancel{E}_T$ and visible transverse momenta, as we will show. 

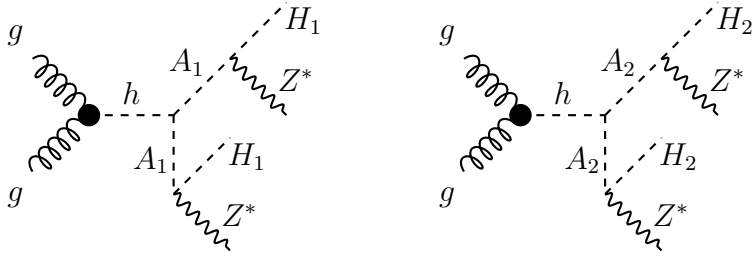
\begin{figure}[t!]
\centering
\begin{tikzpicture}[thick,scale=0.75]
\draw[gluon] (0,0) -- node[black,above,xshift=-0.6cm,yshift=0.4cm] {$g$} (1,-1);
\draw[gluon] (0,-2) -- node[black,above,yshift=-1.0cm,xshift=-0.6cm] {$g$} (1,-1);
\draw[dashed] (1,-1) -- node[black,above,xshift=0.0cm,yshift=0.0cm] {$h$} (2.5,-1);
\draw[dashed] (2.5,-1) -- node[black,above,xshift=-0.2cm,yshift=0cm] {$A_1$} (3.5,0);
\draw[photon] (3.5,0) -- node[black,above,xshift=0.5cm,yshift=-0.2cm] {$Z^*$} (4.5,-1);
\draw[dashed] (3.5,0) -- node[black,above,xshift=0.6cm,yshift=-0.2cm] {$H_1$} (4.5,1);
%%%
\draw[dashed] (2.5,-1) -- node[black,above,yshift=-0.4cm,xshift=-0.3cm] {$A_1$} (2.5,-2.4);
\draw[dashed] (2.5,-2.4) -- node[black,above,xshift=0.6cm,yshift=-0.2cm] {$H_1$} (3.5,-1.4);
\draw[photon] (2.5,-2.4) -- node[black,above,xshift=0.5cm,yshift=-0.2cm] {$Z^*$} (3.5,-3.4);
\draw[xshift=-0cm] (1,-1) node[circle,fill,inner sep=3pt](A){} -- (1,-1);
\end{tikzpicture}
\hspace{10mm}
\begin{tikzpicture}[thick,scale=0.75]
\draw[gluon] (0,0) -- node[black,above,xshift=-0.6cm,yshift=0.4cm] {$g$} (1,-1);
\draw[gluon] (0,-2) -- node[black,above,yshift=-1.0cm,xshift=-0.6cm] {$g$} (1,-1);
\draw[dashed] (1,-1) -- node[black,above,xshift=0.0cm,yshift=0.0cm] {$h$} (2.5,-1);
\draw[dashed] (2.5,-1) -- node[black,above,xshift=-0.2cm,yshift=0cm] {$A_2$} (3.5,0);
\draw[photon] (3.5,0) -- node[black,above,xshift=0.5cm,yshift=-0.2cm] {$Z^*$} (4.5,-1);
\draw[dashed] (3.5,0) -- node[black,above,xshift=0.6cm,yshift=-0.2cm] {$H_2$} (4.5,1);
%%%
\draw[dashed] (2.5,-1) -- node[black,above,yshift=-0.4cm,xshift=-0.3cm] {$A_2$} (2.5,-2.4);
\draw[dashed] (2.5,-2.4) -- node[black,above,xshift=0.6cm,yshift=-0.2cm] {$H_2$} (3.5,-1.4);
\draw[photon] (2.5,-2.4) -- node[black,above,xshift=0.5cm,yshift=-0.2cm] {$Z^*$} (3.5,-3.4);
\draw[xshift=-0cm] (1,-1) node[circle,fill,inner sep=3pt](A){} -- (1,-1);
\end{tikzpicture}
\caption{Diagrams leading to the $4\ell + \Et$ final state mediated by the $h$ boson.}
\label{fig:Et4l}
\end{figure}

Two candidates for  relevant processes here would be $gg \to h \to A_{1}A_{1} \to Z^* H_{1} \, Z^* H_{1}$ and $gg \to h \to A_{2}A_{2} \to Z^* H_{2} \, Z^* H_{2}$, whose final states appear as $4\ell + \cancel{E}_T$,  as shown in Fig.~\ref{fig:Et4l}. 
Since the mass splitting of the two DM candidates, $m_{H_2}-m_{H_1}$, is larger than the $\cancel{E}_T$ resolution at the LHC, one could potentially see the effect of the two DM components in different distributions. 
These channels are particularly interesting since they are sensitive to the $hA_1A_1$ and $hA_2A_2$ couplings, which highlight an important characteristic of the model, i.e., the fact that, due to the sub-dominant relic density of $H_2$, the couplings of the heavier DM family to the SM-like Higgs boson could be exceptionally large ($g_{hA_2A_2} \sim 0.5$) while still in agreement with all collider and astrophysical bounds. 
However, the cross-sections for these processes are very small ($\sim \, 10^{-7}$ fb for the first family and $\sim \, 10^{-4}$ fb for the second family) to have any visible effect in collider searches. As a result, we do not discuss this process any further.

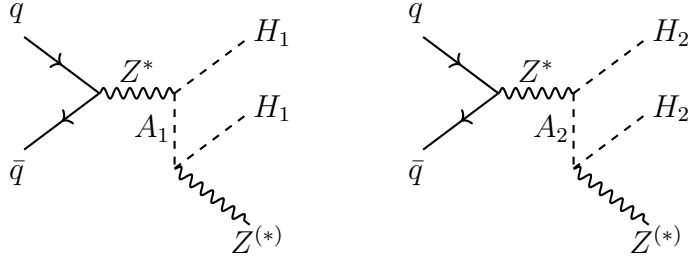
\begin{figure}[t!]
\centering 
\begin{tikzpicture}[thick,scale=1.0]
\draw[particle] (10,0) -- node[black,above,xshift=-0.6cm,yshift=0.4cm] {$q$} (11,-0.75);
\draw[antiparticle] (10,-1.5) -- node[black,above,yshift=-1.0cm,xshift=-0.6cm] {$\bar{q}$} (11,-0.75);
\draw[photon] (11,-0.75) -- node[black,above,xshift=0.0cm,yshift=0.0cm] {$Z^*$} (12,-0.75);
\draw[dashed] (12,-0.75) -- node[black,above,yshift=0.1cm,xshift=0.8cm] {$H_1$} (13,-0);
\draw[dashed] (12,-0.75) -- node[black,above,yshift=-0.3cm,xshift=-0.3cm] {$A_{1}$} (12,-1.75);
\draw[dashed] (12,-1.75) -- node[black,above,yshift=0.1cm,xshift=0.8cm] {$H_1$} (13,-1);
%\draw[photon] (12,-1.75) -- node[black,above,xshift=-0.3cm,yshift=-0.3cm] {$Z^*$} (12,-2.75);
%\draw[particle] (12,-2.75) -- node[black,above,xshift=0.7cm,yshift=-0.1cm] {$l$} (13,-2);
%\draw[antiparticle] (12,-2.75) -- node[black,above,xshift=0.6cm,yshift=-0.6cm] {$\bar{l}$} (13,-3.5);
\draw[photon] (12,-1.75) -- node[black,above,xshift=0.6cm,yshift=-0.9cm] {$Z^{(*)}$} (13,-2.5);
%\node at (11,-4) {(A)};
\end{tikzpicture}
\hspace{1cm}
\begin{tikzpicture}[thick,scale=1.0]
\draw[particle] (10,0) -- node[black,above,xshift=-0.6cm,yshift=0.4cm] {$q$} (11,-0.75);
\draw[antiparticle] (10,-1.5) -- node[black,above,yshift=-1.0cm,xshift=-0.6cm] {$\bar{q}$} (11,-0.75);
\draw[photon] (11,-0.75) -- node[black,above,xshift=0.0cm,yshift=0.0cm] {$Z^*$} (12,-0.75);
\draw[dashed] (12,-0.75) -- node[black,above,yshift=0.1cm,xshift=0.8cm] {$H_2$} (13,-0);
\draw[dashed] (12,-0.75) -- node[black,above,yshift=-0.3cm,xshift=-0.3cm] {$A_{2}$} (12,-1.75);
\draw[dashed] (12,-1.75) -- node[black,above,yshift=0.1cm,xshift=0.8cm] {$H_2$} (13,-1);
\draw[photon] (12,-1.75) -- node[black,above,xshift=0.6cm,yshift=-0.9cm] {$Z^{(*)}$} (13,-2.5);
%\node at (11,-4) {(B)};
\end{tikzpicture}
\caption{Diagrams leading to the $2\ell + \Et$ final state mediated by the $Z$ boson.}
\label{fig:Et2l}
\end{figure}

The ideal processes for our analysis are the $q \bar{q} \to Z^* \to H_1 A_1 \to H_1 H_1 Z^*$ and $q \bar{q} \to Z^* \to H_2 A_2 \to H_2 H_2 Z^*$ ones, 
whose final states appear as $2\ell + \cancel{E}_T$, as shown in Fig.~\ref{fig:Et2l}.
To highlight these signatures, we have chosen four BPs as shown in Tab.~\ref{tab:BPs2}. The cross-sections for the $pp\to 2H_1 + 2\ell$ and $pp\to 2H_2 + 2\ell$ processes for the four BPs are shown in Tab.~\ref{tab:Et2l-xsec}.

\begin{table} [h!]
\begin{center}
\begin{footnotesize}
\begin{tabular}{|c||c|c|c|c|c|c|c|c|c|c|} \hline\\[-2.7ex]
BP & 
$m_{A_1}$ &
$m_{H_1^\pm}$ &
$m_{A_2}$ &
$m_{H_2^\pm}$ &
$\Lambda_2$ &
$g_{hA_1A_1}$ &
$g_{hH_1^+H_1^-}$ &
$g_{hH_2H_2}$   & 
$g_{hA_2A_2}$ &
$g_{hH_2^+H_2^-}$  \\
\hline
L2n2 & 
59.3 &
104.86 &
130.06 &
123.53 &
-0.0053 & 
0.0336141 &
0.280793 &
$-0.0106$   & 
0.217955 &
0.163227 \\
\hline
L2p1 & 
59.3 &
94.6 &
129.7 &
141 &
0.073 & 
0.0336141 &
0.213159 &
0.146   & 
0.371464 &
0.472558 \\
\hline
L2p2 & 
59.06 &
94.6 &
149.07 &
108.96 &
0.065 & 
0.0326753 &
0.213159 &
0.13  & 
0.533922 &
0.191877  \\
\hline
L2p3 & 
59.18 &
97.78 &
143.66 &
140.82 &
0.0099 & 
0.0331442 &
0.233377 &
0.0198   & 
0.371383 &
0.344682 \\
\hline
\end{tabular}
\end{footnotesize}
\caption{\footnotesize The four BPs, for which we have chosen $m_{H_1} = 50$ GeV, $m_{H_2} = 1000$ GeV and $g_{hH_1H_1}=0.00002$, and we have set $\lambda_{11}=0.11$, $\lambda_{22}=0.12$, $\lambda_{12}=0.121$, $\lambda'_{12}=0.13$, $\Lambda_1 = \Lambda_3 = 0.00001$, the SM Higgs mass $m_{h}=125$ GeV and the VEV $v=246$ GeV, are in agreement with all astrophysical and collider constraints.}
\label{tab:BPs2}
\end{center}
\end{table}

We now proceed to a Monte Carlo (MC) analysis of these BPs.

\section{Numerical results}
\label{sec:results}

The integrated and differential cross-sections have been calculated using \texttt{MadGraph} \cite{Alwall:2014hca} by adopting  a generic LHC parameter card.
Furthermore, we have used \texttt{MadAnalysis} \cite{Conte:2012fm}  for constructing our selection, which  involve  basic cuts for leptons $\ell$ and jets 
$j$\footnote{The latter being produced from QCD Initial State Radiation (ISR) and clustered with  the Cambridge/Aachen algorithm with a 0.4 cone size \cite{Dokshitzer:1997in,Wobisch:1998wt}.}: in pseudorapidity $|\eta (\ell) |,  |\eta (j) | < 3$,  transverse momentum $p_T (\ell ), p_T(j) > 10$ GeV as well as separation $\Delta R (j, \ell) >0.5$.  Finally,  event rates have been computed considering $100$ fb$^{-1}$ of  luminosity for the LHC machine.

The cross-sections for the two processes $q \bar{q} \to Z^* \to H_1 A_1 \to H_1 H_1 Z^*$ and $q \bar{q} \to Z^* \to H_2 A_2 \to H_2 H_2 Z^*$ given  in Tab.~\ref{tab:Et2l-xsec} are instrumental to enable the extraction of distributions sensitive to the simultaneous presence of the two DM candidates at the forthcoming
LHC Run 3 as they are both visible and comparable to each other. 
As mentioned before, in all four BPs presented herein,  within each family the cross-section depends on both the absolute value of the masses involved and their mass splittings $M_{A_i}-M_{H_i}$ ($i = 1, 2$), with the latter having been chosen (in agreement with all astrophysical bounds) to be larger than or comparable to the expected experimental resolutions in (missing) transverse energy/momentum and (transverse  or invariant) mass. Therefore, we shall be looking for differential spectra with a distinctive shape from which the existence of  two different underpinning component distributions could be easily inferred.
As a proof of concept, we show such observables for both the aforementioned two processes limitedly to one BP, e.g.,  L2n2. (Results are qualitatively 
similar for the other BPs, so we do not discuss these here.) 
For this choice, it is worth highlighting that the inert masses involved are as follows:  $m_{H_1}=50$ GeV with $m_{A_1}-m_{H_1} =9.3$ GeV for one  doublet and $m_{H_2}=100$ GeV with $m_{A_2}-m_{H_2} =30.06$ GeV for the other.

\begin{table}[t!]
\begin{center}
\begin{footnotesize}
\begin{tabular}{|c|c|c|}
\hline
BP & 
$p p \to 2H_1+ 2 \ell$ &  $p p \to 2H_2+ 2 \ell$ \\
\hline
L2n2&$ \sigma = 1.058 $ [fb]&  $ \sigma = 1.929 $ [fb]\\
\hline
L2p1&$ \sigma = 1.043 $ [fb]& $ \sigma = 1.912 $ [fb]\\
\hline
L2p2 &$ \sigma = 0.9169 $ [fb]& $ \sigma = 1.230 $ [fb]\\
\hline
L2p3 &$ \sigma = 0.9797 $ [fb]&$ \sigma = 2.448 $ [fb]\\
\hline
\end{tabular}
\end{footnotesize}
\caption{The cross-sections for the processes yielding $2\ell + \cancel{E}_T$ final states, whose leading contribution come from the 
$q \bar{q} \to Z^* \to H_1 A_1 \to H_1 H_1 Z^*$ and $q \bar{q} \to Z^* \to H_2 A_2 \to H_2 H_2 Z^*$ channels shown in Fig.~\ref{fig:Et2l}.} 
\label{tab:Et2l-xsec}
\end{center}
\end{table}%

\begin{figure}[h!]
\begin{center}
\includegraphics[scale=0.3]{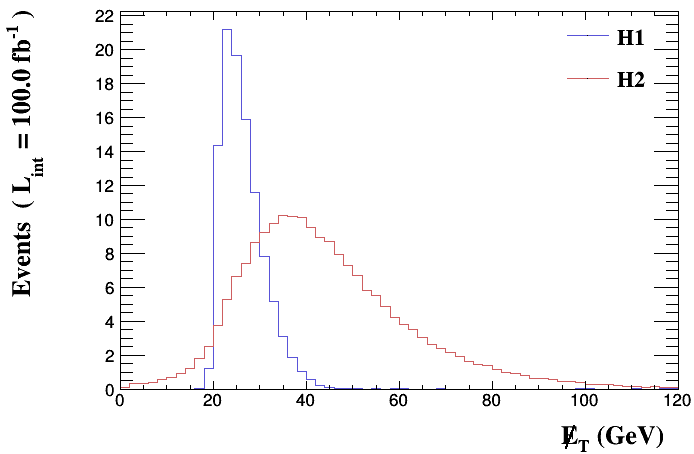}~~~
\includegraphics[scale=0.3]{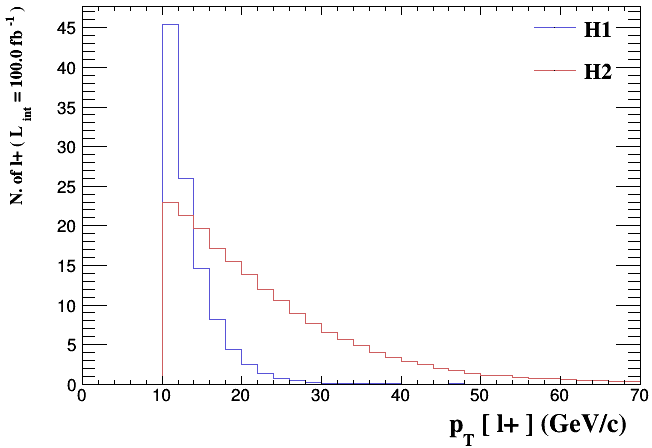}
\caption{Missing transverse energy (left) and transverse momentum of either  lepton (right) in the case of  $H_1$ and $H_2$ separately.}
\label{fig:MET-PT}
\end{center}
\end{figure}

In Fig.~\ref{fig:MET-PT}, we show the $ \cancel{E}_T$ spectrum on the left and that of the transverse momentum of either lepton on the right, for the two channels  
$p p \to \ell \bar{\ell} + H_1H_1$ 
and  $p p \to \ell \bar{\ell}+ H_2H_2$ separately. The two sets of distributions are strikingly different so that, even when summed, one could clearly deduce the
presence of two different DM candidates with different masses. In the $ \cancel{E}_T$ histogram, one can see two different peaks, a sharp one around 25 GeV for the $H_1$ case 
and a smoother one around 35 GeV for the $H_2$ case, this tracking the fact that $m_{H_1}\ll m_{H_2}$ (even though the presence of two DM particles in the event in each case spoils somewhat the correlations). In the 
$p_T^\ell $ spectra, it can be noticed the much sharper decrease of the $H_1$ distribution with respect to the $H_2$ one, this correlating to the fact that $m_{A_1}-m_{H_1} \ll m_{A_2}-m_{H_2} $. 

\begin{figure}[h!]
\begin{center}
\includegraphics[scale=0.3]{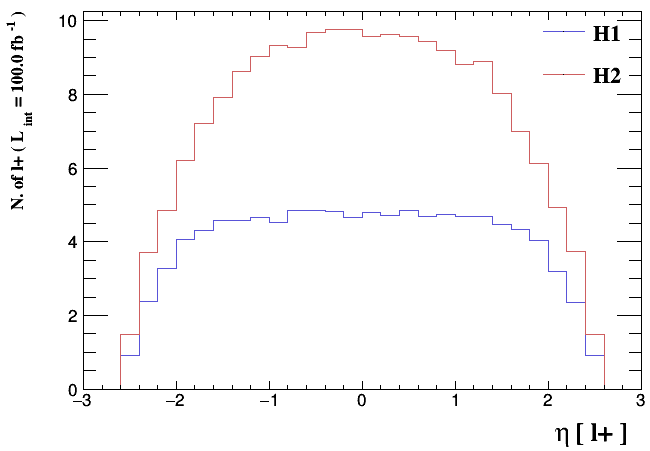}~~~
\includegraphics[scale=0.3]{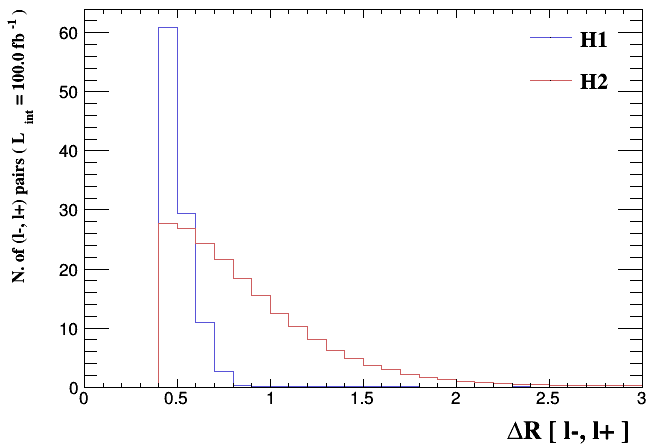}
\caption{Pseudorapidity of either lepton (left) and separation between the leptons (right)  in the case of  $H_1$ and $H_2$ separately.}
\label{fig:eta-deltar}
\end{center}
\end{figure}

In Fig.~\ref{fig:eta-deltar}, one can notice that leptons are similarly central (see the pseudorapidity plot of either of these on the left) while their separation 
is noticeably different (see their cone size on the right) between the two DM candidates, the former spectrum showing that the overall events are not generally boosted but  the $Z^*$ itself can be so and differently between the $H_1$ and $H_2$ cases.

\begin{figure}[h!]
\begin{center}
\includegraphics[scale=0.35]{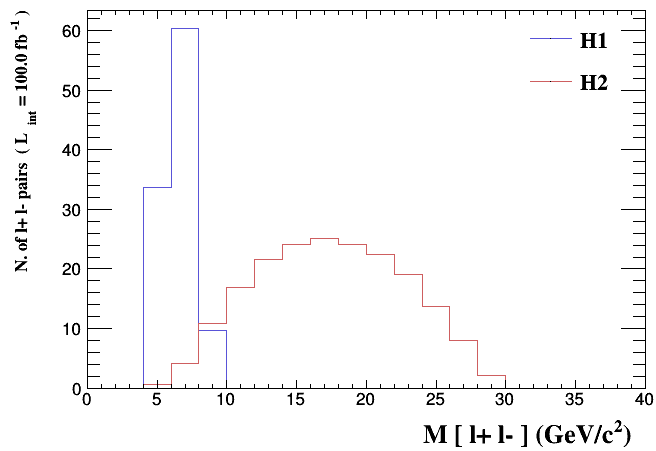}
\caption{The invariant mass of the final state leptons  in the case of  $H_1$ and $H_2$ separately.}
\label{fig:Inv-mass}
\end{center}
\end{figure}

Fig.~\ref{fig:Inv-mass} shows the invariant mass of the leptons, defined through $M_{\ell\ell}^2=(p_{\ell^+}+p_{\ell^-})^2$,  which is also correlated with the $A_i$ and $H_i$ mass splitting in each process, thus 
explaining the low mass peak for the $H_1$ case and the high mass peak for the $H_2$ case.

\begin{figure}[h!]
\begin{center}
\includegraphics[scale=0.35]{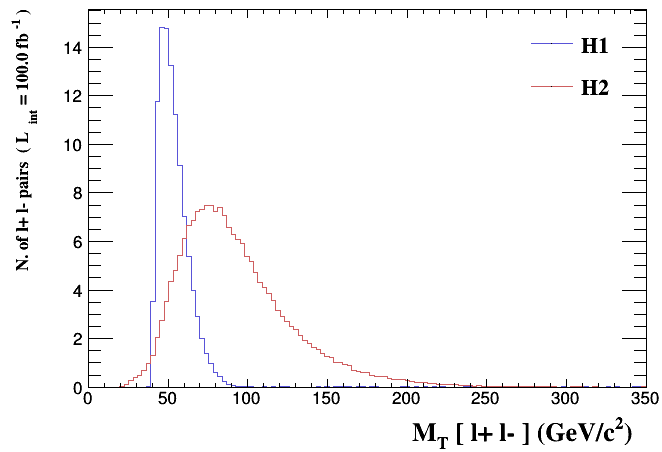}
\caption{The transverse mass of the final state leptons and DM candidates  in the case of  $H_1$ and $H_2$ separately.}
\label{fig:mT}
\end{center}
\end{figure}

Fig.~\ref{fig:mT} shows the transverse mass of the final state leptons, defined through $M_T^2(\ell \ell) = (\sum_i  E_{Ti})^2 - (\sum_i p_{Ti})^2$ , which
is rather strongly correlated to  the $A_1$ and $A_2$ masses for the $H_1$ and $H_2$ cases, respectively.

Therefore, there exist several distributions which are significantly different in shape between the two DM candidates while also yielding sizeable event rates for 
LHC Run 3 luminosities, so that one could not only establish the presence of $H_1$ and $H_2$ simultaneously but also attempt to extract the underlying inert mass spectra from fitting the ensuing data to the I(2+1)HDM predictions. All this is clearly subject to validation through a MC analysis in presence of both reducible and
irreducible backgrounds from  the SM.

\section{Conclusions}
\label{sec:conclusion}

In this paper, we have studied a I(2+1)HDM framework symmetric under a $Z_2\times Z'_2$ group with one inert doublet being odd under $Z_2$ and even under  $Z'_2$ and the other inert doublet being even under $Z_2$ and odd under $Z'_2$, while all SM particles transform trivially under the $Z_2\times Z'_2$ symmetry.
The lightest particle from each inert doublet is a viable DM candidate, resulting in a two-component DM model.
In a recent publication \cite{Hernandez-Sanchez:2020aop}, we showed that, when there is a sufficient mass difference between the two DM candidates, which are both typically at the EW scale, the light DM component can be probed by the nuclear recoil energy in direct detection experiments while the heavy DM  component appears through its contribution to the photon flux in indirect detection experiments. 

Here, in addition to such astrophysical probes, we have shown that certain collider signatures can serve the same purpose; in each family, the decay of the heavier neutral inert state into the lighter one (i.e., the  DM candidate) plus an off-shell (secondary) $Z$ boson (decaying to di-lepton pairs) in association with an additional identical DM candidate emerging in parallel from primary production via an off-shell (primary) $Z$ boson. 
The smoking-gun signature of this two-component DM scenario is thus $2\ell + \cancel{E}_T$, which can in fact be pursued already at the upcoming Run 3 of the LHC. Herein, one could
study a variety of differential distributions stemming  from this final state that would have a distinctive shape carrying the imprint of two underlying components, each
corresponding to a different DM candidate. Indeed, those having an energy dimension could even be used to extract the underpinning inert state masses involved. 

While these conclusions have been obtained without a signal-to-background analysis, we encourage ATLAS and CMS experimentalists to look into this novel phenomenology, as it would manifest itself in a well-studied final state at  the LHC machine.

\subsubsection*{Acknowledgement}
SM acknowledges support from the STFC Consolidated Grant ST/L000296/1 and is partially financed through the NExT Institute.
DS is supported by the National Science Center, Poland, through the HARMONIA project under contract UMO-2015/18/M/ST2/00518.
VK acknowledges financial support from the Academy of Finland projects ``Particle cosmology and gravitational waves'' No. 320123
and ``Particle cosmology beyond the Standard Model'' No. 310130.
J H-S acknowledges support from SNI-CONACYT, PRODEP-SEP and VIEP-BUAP.
The authors acknowledge the use of the IRIDIS High-Performance Computing Facility and associated support services at the University of Southampton in the completion of this work.

\end{document}